\documentclass{ifacconf}

\usepackage{graphicx}      
\usepackage{natbib}        

\usepackage{epsfig} 
\usepackage{mathptmx} 
\usepackage{amsmath} 
\usepackage{color,soul}
\usepackage{enumerate}
\usepackage{tikz}

\newcommand{\redline}{\raisebox{2pt}{\tikz{\draw[-,red,solid,line width = 1.5pt](0,0) -- (5mm,0);}}}
\newcommand{\blackline}{\raisebox{2pt}{\tikz{\draw[-,black,solid,line width = 1.5pt](0,0) -- (5mm,0);}}}
\newtheorem{procedure}{Procedure}
\begin{document}
\begin{frontmatter}

\title{Improving mechanical ventilation for patient care through repetitive control} 


\author[First,Second]{Joey Reinders} 
\author[First]{Ruben Verkade} 
\author[First]{Bram Hunnekens}
\author[Second,Third]{Nathan van de Wouw}
\author[Second]{Tom Oomen}

\address[First]{DEMCON Advanced Mechatronics, Best, The Netherlands.}
\address[Second]{Department of Mechanical Engineering, Eindhoven University of Technology, Eindhoven, The Netherlands.}
\address[Third]{Department of Civil, Environmental and Geo- Engineering, University
	of Minnesota, Minneapolis, MN 55455 USA.}

\thanks{© 2020 the authors. This work has been accepted to IFAC for publication under a Creative Commons Licence CC-BY-NC-ND}
\begin{abstract}
	Mechanical ventilators sustain life of patients that are unable to breathe (sufficiently) on their own. The aim of this paper is to improve pressure tracking performance of mechanical ventilators for a wide variety of sedated patients. This is achieved by utilizing the repetitive nature of sedated ventilation through repetitive control. A systematic design procedure of a repetitive controller for mechanical ventilation is presented. Thereafter, the controller is implemented in an experimental setup showing superior tracking performance for a variety of patients. 
\end{abstract}


\end{frontmatter}

\section{Introduction} \label{sec:Intro}

Mechanical ventilators are essential equipment in Intensive Care Units (ICUs) to assist patients who cannot breathe on their own or need support to breathe sufficiently. The goal of mechanical ventilation is to ensure adequate oxygenation and carbon dioxide elimination \citep{Warner2013}, and thereby sustaining the patient's life. In 2005 over 790,000 patients required ventilation in the United States alone \citep{Wunsch2010}. Therefore, improving mechanical ventilation improves treatment for a large population worldwide.

\begin{figure}[t!]	
	\centering
	\includegraphics[width=0.8\columnwidth]{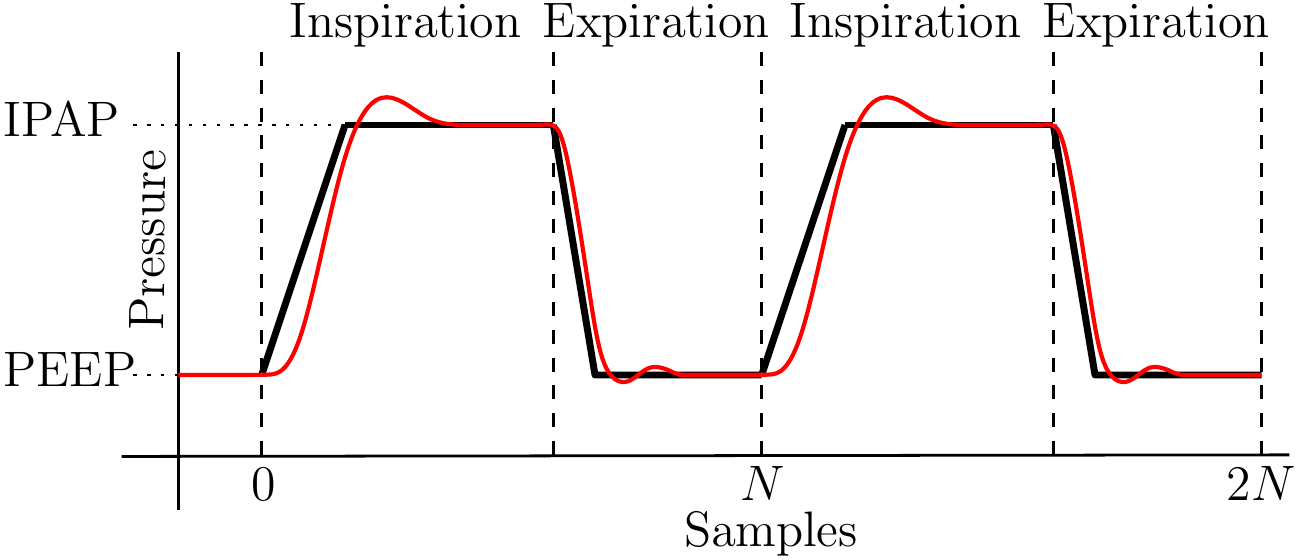}	\caption{Typical airway pressure for two breathing cycles of pressure controlled ventilation, showing the set-point (\protect \blackline) and the typical response (\protect \redline).} 
	\label{fig:BreathingCycle}
\end{figure}

In pressure controlled ventilation modes, the mechanical ventilator aims to track a clinician set pressure profile at the patient's airway. An example of such profile is depicted in Fig. \ref{fig:BreathingCycle}. The Inspiratory Positive Airway Pressure (IPAP) and Positive End-Expiratory Pressure (PEEP) induce airflow in and out of the lungs, respectively. This alternating flow of air allows the lungs to exchange CO$_2$ for O$_2$ in the blood.

Accurate tracking of the preset pressure profile is essential to ensure sufficient patient support and has spurred substantial research to improve control performance. According to \cite{Hunnekens2020}, improved pressure tracking can prevent patient-ventilator asynchrony. In \cite{Blanch2015}, patient-ventilator asynchrony is even associated with increased mortality rates. Furthermore, accurate tracking for a wide variety of patients improves consistency of treatment over these different patients. 

The challenging problem of pressure tracking in presence of uncertain patients has spurred the development of a wide range of pressure control methodologies. In \cite{Hunnekens2020}, variable-gain control is applied to overcome the trade-off between fast rise times and small overshoot. A significant improvement in tracking performance is shown in this work. However, the tracking error is still significant and the patient flow is used in the control strategy, which is typically not available. In \cite{Borrello2001}, adaptive feedback control is used. A patient model is estimated and the controller is adaptively tuned to obtain the desired transfer-function characteristics. It is shown to significantly improve performance in an experimental setting. However, in practice, i.e., on actual patients, it is complex to obtain an accurate patient model, therewith performance is expected to deteriorate.  In \cite{Scheel2017}, a model-based control approach is applied to mechanical ventilation. It is shown that this can improve performance. However, an accurate patient model is required which are typically not available in practical scenarios. In \cite{Li2012}, model predictive control is used to improve tracking performance in ventilation. However, this method also requires an accurate patient which is typically not available. In \cite{Reinders2020}, adaptive hose-compensation control is used to significantly improve tracking performance in ventilation. However, using the repetitive nature of breathing, tracking performance can be improved even further.

Iterative Learning Control (ILC) and Repetitive Control (RC) can achieve superior tracking performance utilizing limited model information and the repetitive nature of a disturbance, e.g., the reference.  ILC (\cite{Arimoto1984}, \cite{Bristow2006}, and \cite{Moore1993}) and RC (\cite{Hara1988}, \cite{Inoue1981}, \cite{Longman2010}, and \cite{Pipeleers2009}) are well-known control strategies. In these methods, the controller learns an input signal using errors of previous tasks. In other application fields, ILC and RC are extensively analyzed and successfully implemented, e.g., industrial robotics \citep{Arimoto1984}, wafer stages \citep{Roover2000}, printer systems (\cite{Zundert2016} and \cite{Blanken2019b}), and in medical applications for a device to help with stroke rehabilitation \citep{Freeman2012}.

Since an exact patient model is typically unavailable, such learning methods are particularly suitable for mechanical ventilation. In \cite{Scheel2015} and \cite{Castro2019}, ILC has been applied to mechanical ventilation. In \cite{Castro2019}, a significant improvement in pressure tracking performance is shown. However, only simulation results are presented. In \cite{Scheel2015}, a strong improvement in tracking performance is shown in experiments. However, only causal filters are used in the ILC design. In sharp contrast, non-causal filters could potentially improve performance significantly because of the delays in ventilation systems \citep{Borrello2005}. Furthermore, in this paper, it is argued that RC is a more suitable approach in a mechanical ventilation setting, because it is a continuous process, i.e., the system states are not reset between repetitions.

Although control has substantially improved tracking performance of ventilation, data of previous breaths and a learning control approach can be used to improve tracking performance by compensating all repeating disturbances. Therefore, the aim of this paper is to improve pressure tracking performance for fully sedated patients utilizing the periodic nature of their breaths.

The main contribution of this paper is the design of an RC in application to mechanical ventilation that achieves superior tracking performance for a wide variety of patients. A step-by-step design process of this RC is presented. Thereafter, this controller is implemented in an experimental setup and the performance is analyzed for a variety of plants, varying from a baby to an adult patient. 

The outline of this paper is as follows. In Section \ref{sec:ConProb}, the  control problem and envisioned solution are presented. Thereafter, in Section \ref{sec:RC}, the control concept, stability results, and design procedure for RC are briefly explained. Then, in  Section \ref{sec:Applied}, the design process of the RC for mechanical ventilation is explained in detail and performance of the controller is analyzed in experiments. Finally, in Section \ref{sec:Conc}, the main conclusions are presented.

\section{Control problem}\label{sec:ConProb}
In this section, the control problem is described in detail. In Section \ref{subsec:SystemDescription}, a high-level system description of the considered ventilation setup is given. Then, in Section \ref{subsec:ControlProblemChallenges}, the control problem and main challenges are described. Finally, the envisioned control approach is briefly described in Section \ref{subsec:EnvisionedApproach}. 

\subsection{System description}\label{subsec:SystemDescription}
\begin{figure}[t]
	\includegraphics[width=0.98\columnwidth]{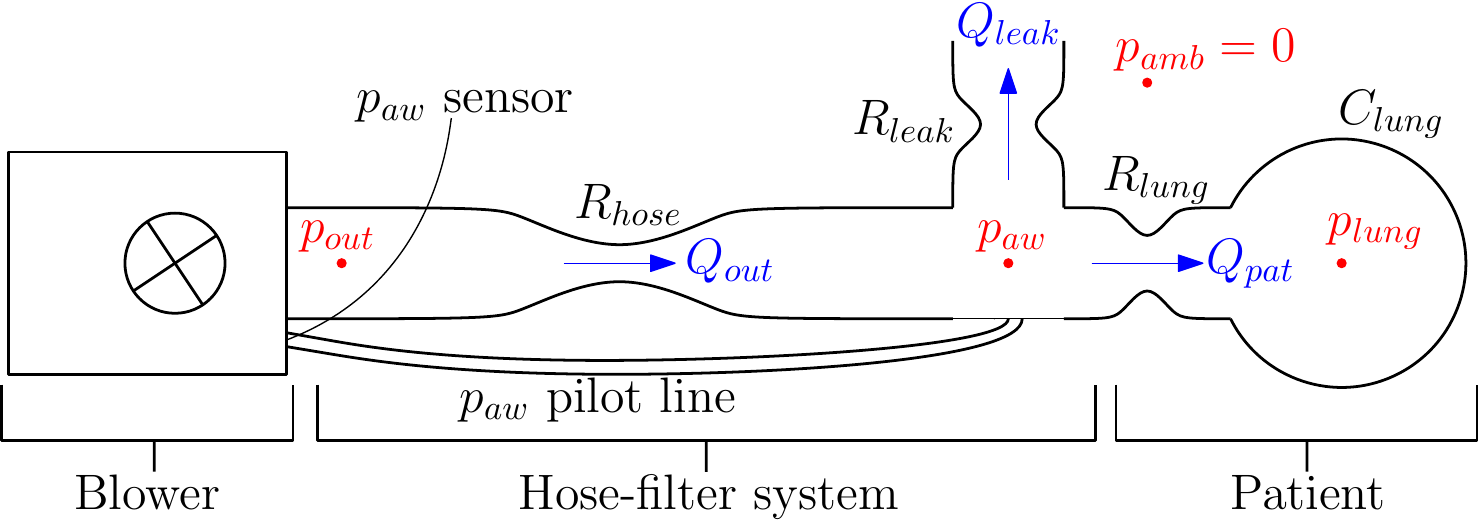}
	\caption{Schematic representation of the blower-hose-patient system, with the corresponding resistances, lung compliance, pressures, and flows.}
	\label{fig:DetailedModelDiscription}
\end{figure}

A schematic of the considered blower-patient-hose system, with the relevant parameters, is shown in Fig. \ref{fig:DetailedModelDiscription}. The main components in the system are the blower, the hose-filter system, and the patient. 

A centrifugal blower compresses ambient air to achieve the desired blower outlet pressure $p_{out}$. This change in $p_{out}$ is controlled to achieve the desired airway pressure $p_{aw}$ near the patient's mouth. The airway pressure is measured using a pilot line attached to the module and the end of the hose. All pressures are defined relative to the ambient pressure, i.e., $p_{amb}$ = 0.

The hose-filter system connects the blower to the patient. The difference between the outlet pressure and the airway pressure results in a flow through the hose $Q_{out}$, related by a hose resistance $R_{hose}$. The change in airway pressure $p_{aw}$ results in two flows, namely, the leak flow $Q_{leak}$ and the patient flow $Q_{pat}$. The leak flow is used to flush exhaled CO$_2$-rich air from the hose. The patient flow is required to ventilate the patient.

The patient is modeled as a resistance $R_{lung}$ and a compliance $C_{lung}$. The patient flow is a result of the lung resistance and the difference between the airway pressure and the lung pressure $p_{lung}$, i.e., the pressure inside the lungs. The patient flow results in a change in the lung pressure, the relation between patient flow and lung pressure is given by the lung compliance. In this paper, Frequency Response Function (FRF) models of the experimental setup are considered as presented in Section \ref{subsec:LFiltDesign}.

\subsection{Control problem and challenges} \label{subsec:ControlProblemChallenges}
This paper considers Pressure Controlled Mandatory Ventilation (PCMV) of fully sedated patients. The goal in PCMV is to track a given airway pressure reference, i.e., preset by the clinician, repeatedly, see Fig. \ref{fig:BreathingCycle} for an example reference. This reference is exactly periodic with a period length of $N$ samples. In case of a fully sedated patient N is preset by the clinician and exactly known. Besides this reference pressure, no other disturbances are considered to be present.

In mechanical ventilation, it is challenging to track such references accurately because of the large plant variations. One single controller should achieve accurate tracking for all possible plant variations. The following components of the plant are typically unknown and varying between patients:
\begin{itemize}
	\item the patient can vary from a neonate to an adult;
	\item the hose and leak resistance can vary dependent on the available hoses in the hospital;
	\item the exact blower dynamics vary slightly from module to module.
\end{itemize}
Due to these unknown plant variations accurate inverse plant feedforward and high-gain feedback control are infeasible.

Besides the plant variations, another challenge is the presence of delays in the system. In the system a blower delay is present from the control output $p_{control}$ to the blower outlet pressure $p_{out}$. Furthermore, delays from $p_{out}$ to $p_{aw}$ and from the actual airway pressure $p_{aw}$ to the measured airway pressure are present. Such delays typically deteriorate performance of classical feedback control strategies.

Because of the plant variations, the delays in the system, and the repetitive nature of the reference signal, ILC and RC are considered particularly suitable control approaches for this application. These control approaches in application to this control problem are briefly discussed in the following section.

\subsection{Control approach} \label{subsec:EnvisionedApproach}
ILC and RC are control approaches that learn a control signal to suppress the tracking error caused by a reproducing disturbance (or reference). The key aspect in these approaches is that control actions are updated based on measured data from past disturbance realizations. Only limited model knowledge is utilized to guarantee fast and stable convergence. Using limited model knowledge and sufficient data, i.e., repetitions of the reproducing disturbance, the consequences of inevitable modelling errors can be suppressed and tracking performance can be improved. Therewith, these approaches can achieve accurate tracking for wide plant variations using a model of an 'average' system. Furthermore, these control approaches allow non-causal control actions, since the control action is a priori known. This can result in a significant performance gain in systems with delays, such as the considered ventilation system.

The key difference between ILC and RC is that in ILC the system is reset in between tasks, i.e., the initial conditions are exactly the same, whereas in RC the system operates in continuous time, without reset. Mechanical ventilation is a continuous process and there is no reset between breaths. Hence, the initial conditions of a breath depend on the previous breaths and inputs and are not always the same. Therefore, RC is considered a more suitable control approach for the application of mechanical ventilation. In the next section, the concept and theory of RC are treated.
\section{Repetitive control}\label{sec:RC}
\begin{figure}[t]
	\centering
	\includegraphics[trim={0.5cm 0cm 2cm 0cm},clip,width=0.8\columnwidth]{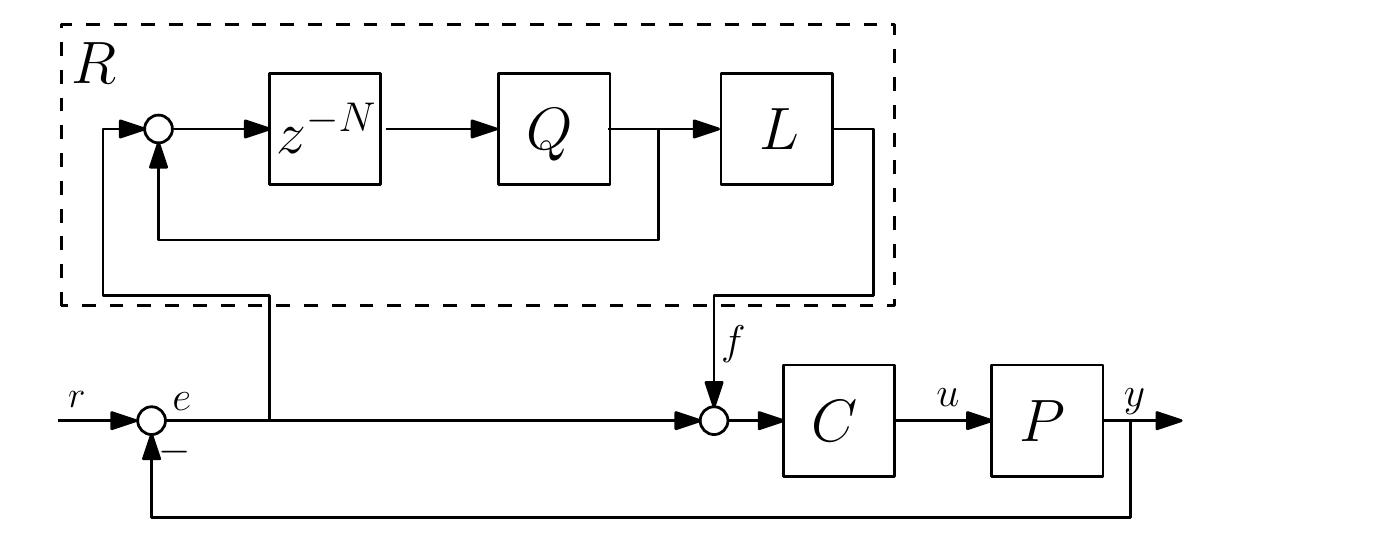}
	\caption{Block diagram of a classic feedback control system including an add-on repetitive controller.}
	\label{fig:BlockSchemeRC}
\end{figure}

A closed-loop control system with a feedback controller and an add-on RC is depicted in Fig. \ref{fig:BlockSchemeRC}. In this figure, $P$ denotes the plant, $C$ is a linear stabilizing feedback controller, $R$ is the add-on RC, the robustness filter is denoted by $Q$, the learning filter is denoted by $L$, and $N$ denotes the length of the reproducing disturbance, i.e., the reference $r$, in samples. The repetitive controller is designed in the $z$-domain, based on a discrete-time plant model $P$ in Fig. \ref{fig:BlockSchemeRC}.

RC is based on the Internal Model Principle (IMP) \cite{Francis1975}. The IMP states that asymptotic disturbance rejection of an exogenous disturbance is achieved if a model of the disturbance generating system is included in a stable feedback loop. For general $N$-periodic disturbances, a model of the disturbance generating system can be obtained using a memory loop. Including this memory loop in the control loop, see Fig. \ref{fig:BlockSchemeRC} with $Q=L=1$, results in a transfer function from the reference to the error with infinite rejection at the harmonics of $N$. Hence, a reference signal that is exactly periodic with period length $N$ is perfectly rejected.

In the remainder of this section, stability and filter design for RC are explained in Section \ref{subsec:StabRC} and \ref{subsec:LQdesRC}, respectively. 

\subsection{Stability analysis of repetitive control} \label{subsec:StabRC} 
In this section, stability properties of the controlled system with an RC are presented. The presented stability conditions are a special case of the conditions in \citet[Theorem 4]{Longman2010}. First, the full closed-loop transfer function is obtained from Fig. \ref{fig:BlockSchemeRC} 
\begin{equation}
\label{eq:error1}
\begin{split}
e &= (I+PC(I+R))^{-1}r \\
&= (I+PC + (I+PC)(I+PC)^{-1}PCR)^{-1}r \\
&= (I+PC + (I+PC)TR)^{-1}r\\
&= ((I+PC)(I+TR))^{-1}r=\underbrace{(I+TR)^{-1}}_\text{$S_R$} \underbrace{(I+PC)^{-1}}_\text{$S$}r,
\end{split}
\end{equation}
where $S$ is the sensitivity of the closed-loop system with $R=0$, $T$ is the complementary sensitivity with $R=0$, i.e., $T = 1-S$, and $S_R$ is referred to as the modifying sensitivity. It is assumed that the sensitivity $S$ is asymptotically stable due to design of $C$. Using this assumption and \eqref{eq:error1}, it follows that the closed-loop is asymptotically stable if and only if $S_R$ is asymptotically stable.\\

By substituting the transfer function of $R:=Lz^{-N}Q( I-z^{-N}Q)^{-1}$ in $S_R$, it is obtained that
\begin{align}
\label{eq: SR filter}
S_R &= (I-z^{-N}Q)(I-(I-TL)z^{-N}Q)^{-1}.
\end{align}

This equation depends on the period length $N$, however, stability properties independent of $N$ are desired. Stability conditions independent of $N$ are desired because the breath length is often changed by a clinician. Hence, conditions independent of $N$ allow for filter design independent of $N$. Therefore, the Single-Input Single-Output (SISO) stability condition in Theorem \ref{th:stabRC} below is commonly used, which is a special case of the multi-variable case in \citet[Theorem 4]{Longman2010} and is independent of $N$. Since this theorem is independent of $N$, controller design can be done independent of the period length $N$.

\begin{thm}	\label{th:stabRC}	
	Assume that $S$ and $T$ are asymptotically stable. Then, $S_R$ is asymptotically stable for all $N$ if
	
	\begin{align}\label{eq:Theorem}
	|Q(z)(1-T(z)L(z))|<1, \forall z=e^{i\omega}, \omega \in [0,2\pi). 
	\end{align}
\end{thm}
Essentially, this theorem ensures that $-(1-TL)z^{-N}Q$ in \eqref{eq: SR filter} stays inside the unit-circle and, therewith, encirclements of the -1 point never occur, i.e., $S_r$ is asymptotically stable. This implies that the entire loop in Fig. \ref{fig:BlockSchemeRC} is asymptotically stable.

\subsection{Filter design for repetitive controller} \label{subsec:LQdesRC}
Using the stability condition in Theorem \ref{th:stabRC}, the following two-step design procedure is followed for SISO RC systems, see \cite{Hara1988}, \cite{Steinbuch2002}, \cite{Tomizuka1989}, and \cite{Blanken2019b}:

\begin{procedure} (Frequency-domain SISO RC design, from \cite{Blanken2019b}). \label{proc:designproc}
	
	\begin{enumerate}	
		
		\item Given a parameteric model of the 'nominal' complementary sensitivity $T(z)$, construct $L(z)$ as an approximate stable inverse of $T(z)$, i.e., $L(z)\approx T^{-1}(z)$.
		\item Using non-parametric FRF models, $T_i(e^{i\omega})$, $i\in\{1,\dots N_p\}$ with $N_p$ the number of patient models, of different patients, design one $Q(z)$ such that Theorem \ref{th:stabRC} is satisfied for $T_i(e^{i\omega})~\forall i\in\{1,\dots N_p\}$.
	\end{enumerate}
\end{procedure}

This procedure describes a systematic robust design method for RC. In step 1, the $L$ filter is based on a coarse parametric model of a 'nominal' patient. In case $T$ is non-minimum phase or strictly proper, algorithms such as Zero Phase Error Tracking Control (ZPETC), see \cite{Tomizuka1987}, can be used to obtain a stable $L$ filter. Then, in step 2, robustness to modeling errors and plant variations can be handled effectively. This is done by using non-parametric FRF models of the complementary sensitivity, which are easily obtained from experimental data \cite{Pintelon2012}. Using FRF models of different plants, stability for these different plants can be ensured using step 2 of the procedure. In the following section, the described design procedure is applied to the considered mechanical ventilation system.

\section{Repetitive control applied to mechanical ventilation scenarios}\label{sec:Applied}
Next, RC is applied to a mechanical ventilation setup. First, the experimental ventilation setup is described in Section \ref{subsec:Setup}. The learning filter and robustness filter are designed in Section \ref{subsec:LFiltDesign} and \ref{subsec:QFiltDesign}, respectively. Finally, the RC is implemented on the experimental setup and performance is compared to PID control in Section \ref{subsec:PerformanceAnalysis}.

\subsection{Setup and use-case description}\label{subsec:Setup}
\begin{figure}[t]
	\centering
	\includegraphics[trim={0cm 0 0cm 0cm},clip,width=0.8\columnwidth]{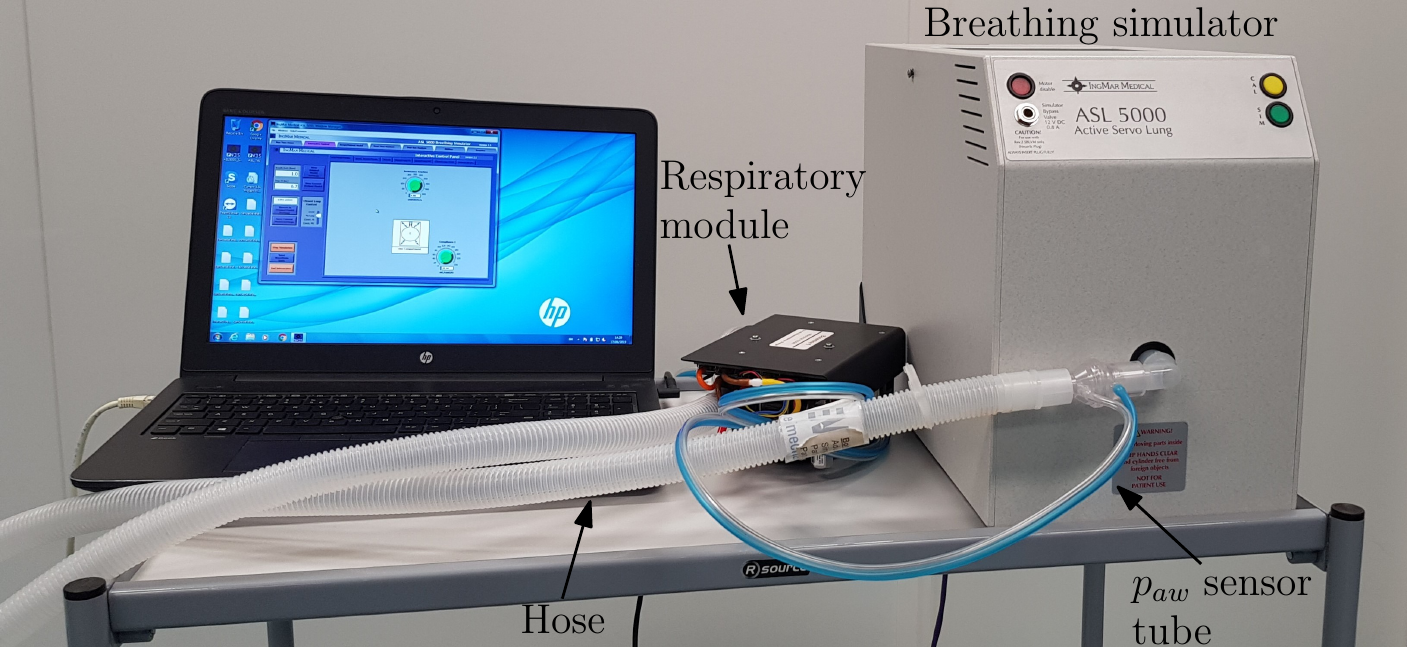}
	\caption{Experimental setup consisting of the blower driven ventilator, ASL 5000 breathing simulation, and a hose.}
	\label{fig:setup}
\end{figure}

The main components of the experimental setup used in this case study are depicted in Fig. \ref{fig:setup}. The figure shows a Macawi blower-driven mechanical ventilation module (DEMCON macawi respiratory systems, Best, The Netherlands). Furthermore, the ASL 5000\texttrademark Breathing Simulator (IngMar Medical, Pittsburgh, PA) is shown in the figure. This breathing simulator is used to emulate a linear one-compartmental patient model as described in \cite{Bates2009}. Furthermore, a typical hose-filter system for ventilation of a patient in a hospital setting is shown. The developed control algorithms are implemented in a dSPACE system (dSPACE GmbH, Paderborn, Germany), which is not shown in the figure.

To design and evaluate an RC for mechanical ventilation, three different patients and ventilation scenarios are considered. The considered patient scenarios are a baby, pediatric, and adult scenario from the ISO standard for PCMV obtained from Table 201.104 in NEN-EN-ISO 80601-2-12:2011 (NEN, Delft, The Netherlands). For these standardize scenarios, the patient parameters and the ventilator settings are given in Table \ref{tab:PatientScenarios}. Note that all scenarios use the same hose-filter-leak configuration. The PID controller that is used in all scenarios, and in the benchmark PID control strategy is a pure integral controller. The benchmark controller is implemented as shown in Fig. \ref{fig:BlockSchemeRC} with $R=0$. This controller is robustly designed to satisfy performance specifications and ensure stability for a large variation of plants. The transfer function of this controller is $C(z) = \frac{0.01257}{z-1}$, with sampling time $2\times10^{-3}$ s. 

\begin{figure}[!tb]
	\centering
	\includegraphics[width=0.8\columnwidth]{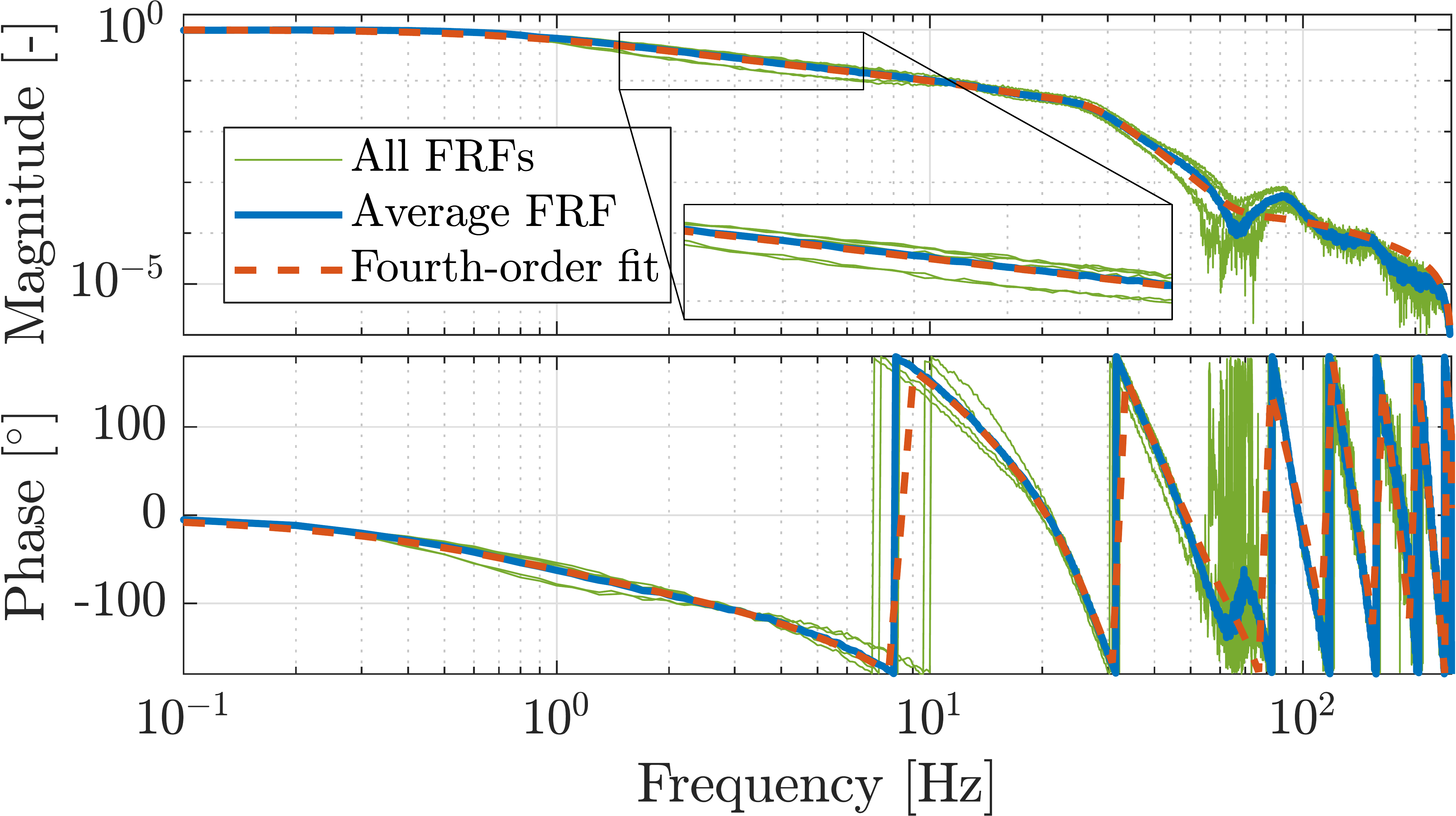}
	\caption{Complementary sensitivity of the different measured FRFs, an average of these FRFs, and a fourth-order fit of the average FRF.}
	\label{fig:FRFMeas}
\end{figure}

To design the RC filters, a Frequency Response Function (FRF) of the complementary sensitivity is identified for every patient scenario at the PEEP pressure level and the IPAP pressure level. These FRFs and the mean of these FRFs are shown in Fig. \ref{fig:FRFMeas}. The mean FRF is used to design the learning filter as described in Section \ref{subsec:LFiltDesign}. To ensure stability, all separate FRFs are used in the design process of the $Q$ filter in Section \ref{subsec:QFiltDesign}.

\begin{table}[t]
	\centering
	\caption{Patient parameters and ventilation settings used for filter design and in the experiments.}	\label{tab:PatientScenarios}
	\begin{tabular}{lllll}
		\textbf{Parameter} & \textbf{Adult} & \textbf{Pediatric} & \textbf{Baby} & \textbf{Unit} \\ \hline
		$R_{lung}$ & 5 & 50 & 50 & mbar s / L \\
		$C_{lung}$ & 50  & 10 & 3 & L/mbar $\cdot 10^{-3}$ \\
		Respiratory rate & 15 & 20 & 30 & breaths / min \\
		PEEP & 5 & 5 & 10 & mbar \\
		IPAP & 15 & 35 & 25 & mbar \\
		Inspiratory time & 1.5  & 1 & 0.6 & s \\
		Expiratory time & 2.5 & 2 & 1.4 & s		
	\end{tabular}
\end{table}

\subsection{$L$-filter design}\label{subsec:LFiltDesign}
According to step one in Procedure \ref{proc:designproc}, the learning filer $L$ should be designed as an approximate stable inverse of the complementary sensitivity $T$. If $L=T^{-1}$ stability is always ensured, independent of the choice for $Q$, see Theorem \ref{th:stabRC}. 

Next, $L$ filter design specifically for this application is considered. FRFs of different patients are displayed in Fig. \ref{fig:FRFMeas}. However, it is desired to design one controller for all patients. Therefore, for the design of $L$, the average FRF in Fig. \ref{fig:FRFMeas} is used for $L$ filter design. There is a significant time delay in the transfer function from the reference to the airway pressure, which is also observable in the FRFs. This time delay $\tau_d$ is identified to be approximately $24$ ms or $12$ samples. The main cause of this delay is the finite propagation speed of pressures waves through the hose and sensor line.

Therefore, a parametric estimate of the mean complementary sensitivity with $24$ ms of delay is obtained. For the estimate of the average plant a fourth-order fit is used, which is shown in Fig. \ref{fig:FRFMeas}; this fit is denoted by $T_{fit}$. $T_{fit}$ is obtained by using the \emph{ssest} function in MATLAB (MathWorks, Natick, MA). This function uses a prediction error minimization \citep{Ljung1999} to estimate a state-space model of the mean plant. It is shown that this fit is accurate up to approximately 30 Hz. This is considered sufficient for this application, since, the target profile contains mainly low frequency information, typically up to 15 Hz. Furthermore, a higher-order fit might improve the fit for these specific patients, but might be less accurate for other patients.

Next, $T_{fit}$ is used to design the $L$ filter. According to step one in Procedure \ref{proc:designproc}, $L$ should be designed as $T_{fit}^{-1}$. A strictly proper system has an improper inverse. Therefore, $T_{fit}$ is inverted using ZPETC. This gives a causal $L$ filter, $L_c$, which is used to obtain a non-causal learning filter
\begin{align}
L = z^{p+d}L_c,
\end{align}
where $p$ and $d$ are the relative degree of $T_{fit}$ and the number of samples delay in $T_{fit}$, respectively. Because of the memory loop, with $N$ samples delay, this non-causal filter can easily be implemented as long as $p+d\leq N$. Next, the design procedure of $Q$ is described to ensure stability of the closed-loop system.

\subsection{$Q$-filter design}\label{subsec:QFiltDesign}
\begin{figure}[t]
	\centering
	\includegraphics[trim=0.0cm 0cm 0.0cm 0cm,clip,width=1\columnwidth]{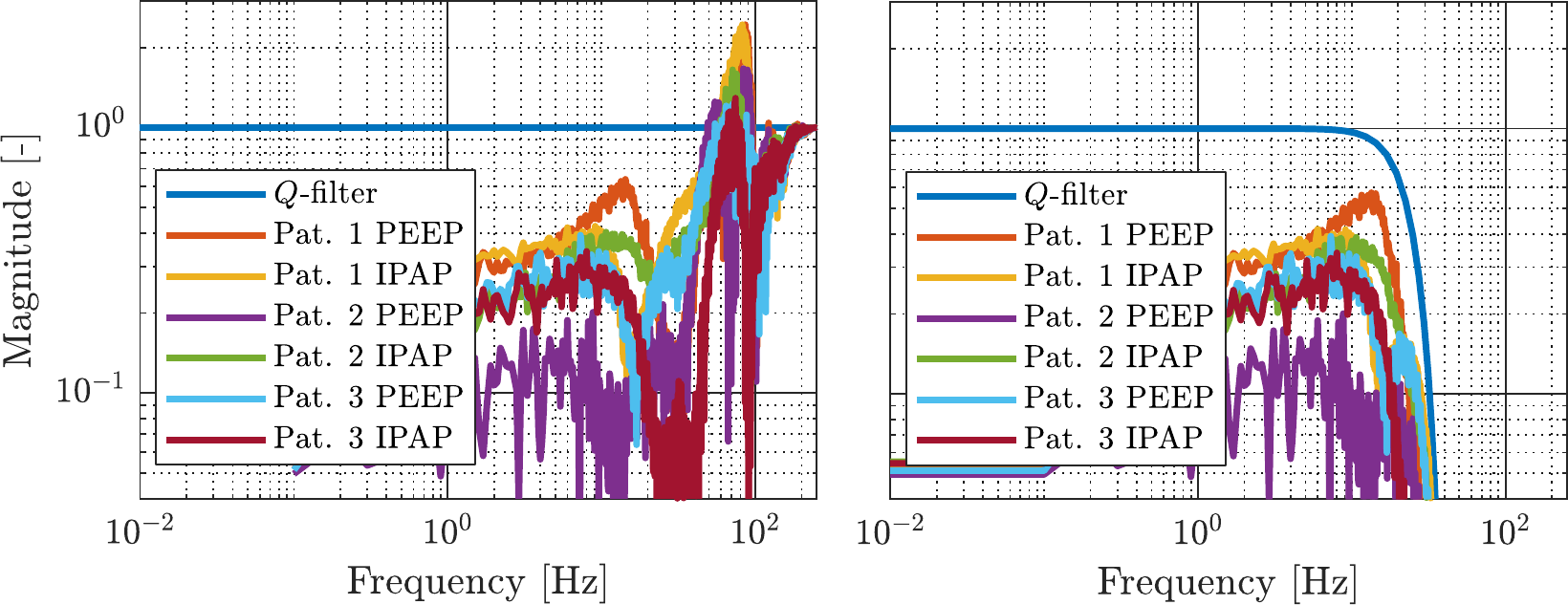}
	\caption{Left: stability condition for all FRFs with $Q=1$. Right: stability conditions for all FRFs with $Q$ a low-pass filter with cut-off frequency at $23$ Hz. The stability criterion in Theorem \ref{th:stabRC} is not ensured for $Q=1$ and stability is guaranteed for all patient with $Q$ a low-pass filter.}
	\label{fig:StabilityFigure}
\end{figure}

Next, step 2 in Procedure \ref{proc:designproc} is followed to ensure stability, i.e., to ensure satisfaction of the conditions in Theorem \ref{th:stabRC}. In particular, a robustness filter $Q$ is included to guarantee closed-loop stability and to improve robustness against plant variations. First, the stability condition in Theorem \ref{th:stabRC} is checked for the system without $Q$ filter, i.e., $Q=1$. This result is shown in the left plot of Fig. \ref{fig:StabilityFigure}, Pat. 1, Pat. 2, and Pat. 3 correspond with the adult, pediatric, and baby patient of Table \ref{tab:PatientScenarios}. Clearly, \eqref{eq:Theorem} does not hold for any patient. That reveals that stability cannot be guaranteed for any patient, see Theorem \ref{th:stabRC}.

Because implementing just an $L$ filter does not guarantee stability, a $Q$ filter is included. This $Q$ filter is designed such that the stability condition in Theorem \ref{th:stabRC} is ensured for all FRFs in Fig. \ref{fig:FRFMeas}. To ensure the stability condition, a low-pass filter with cut-off frequency of 23 Hz is used. This filter is implemented as a 50th order non-causal zero-phase Finite Impulse Response (FIR) filter. This FIR filter is implemented by computing a causal symmetric FIR-filter $Q_c$ and applying a forward shift of $z^{p_q}$ with $p_q$ half the order of the FIR-filter. This makes it a zero-phase FIR filter that is symmetric around zero lag, such that no phase lag is introduced by the filter. The forward shift is possible because of the memory loop, as long as $p+d+p_q \leq N$.

Implementing the designed $Q$ filter ensures the stability condition of Theorem \ref{th:stabRC}, see the right plot in Fig. \ref{fig:StabilityFigure}. Note that the cut-off frequency could be slightly higher, while still ensuring closed-loop stability. However, some margin between $|Q(1-T_{FRF}L)|$ and $1$ is desired to improve robustness against plant variations. Next, the performance of the system with and without RC is compared for all three patient scenarios.

%

\subsection{Performance analysis}\label{subsec:PerformanceAnalysis}
In this section, the performance of the benchmark controller is compared to the performance of the developed RC. This comparison is executed for the different patient scenarios which are given in Table \ref{tab:PatientScenarios}. First, the results of the adult scenario are thoroughly analyzed. Thereafter, the main results of the other two scenarios are briefly presented.

\begin{figure}[t]
		\centering
		\hspace{0cm}\includegraphics[trim=0.0cm 0cm 0.0cm 0cm,clip,width = 0.95\columnwidth]{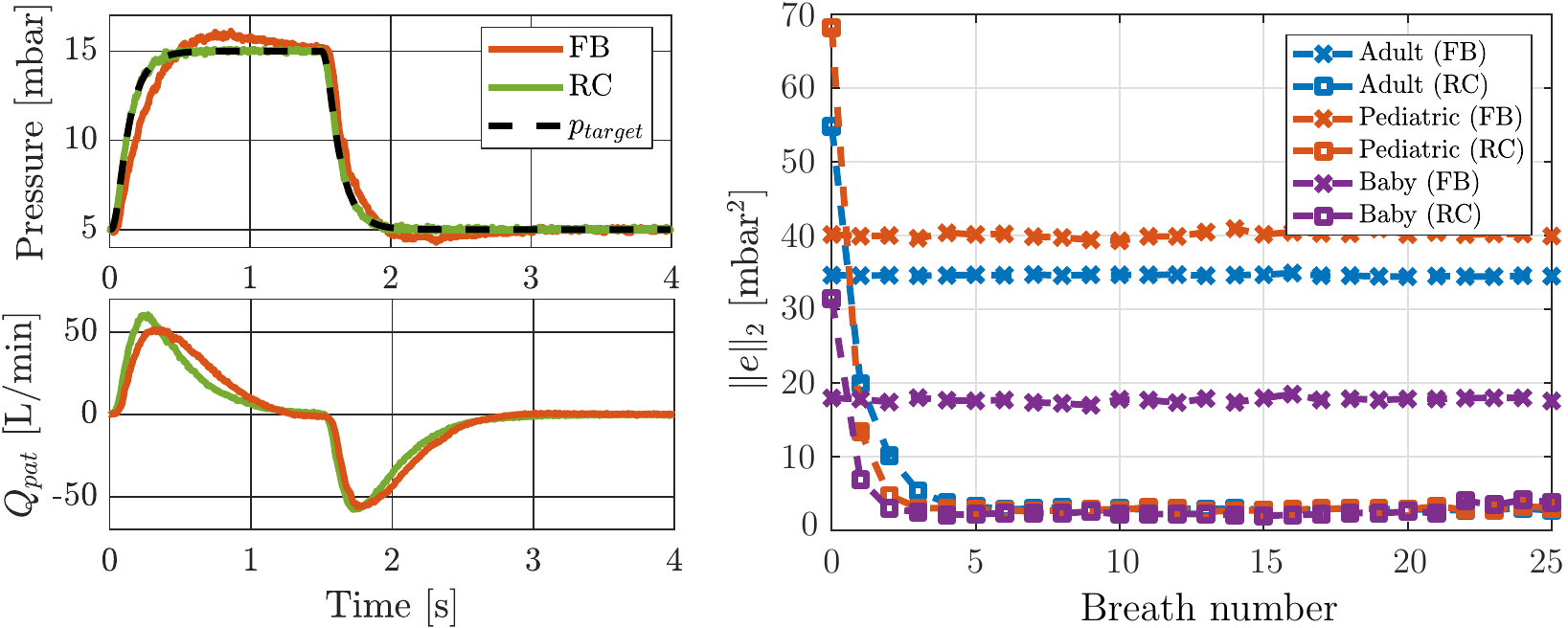}
		\caption{Left: airway pressure $p_{aw}$ and patient flow $Q_{pat}$ of adult scenario (breath 20). Right: error 2-norm of all cases comparing PID with RC.}
		\label{fig:pawenorm}
	\end{figure}

The results of the experiments are given in Fig. \ref{fig:pawenorm}. The left-hand figure shows the airway pressure and the patient flow of the 20th breath of the adult scenario. The right-hand figure shows the error 2-norm of both control strategies per breath, where the tracking error is defined as
$e(t)=p_{target}(t)-p_{aw}(t)$.

Figure \ref{fig:pawenorm} shows that the benchmark controller has significant overshoot and undershoot, and a much longer settling time. Furthermore, it shows that the RC makes the airway pressure almost exactly the same as the target pressure upon convergence. Hence, pressure tracking performance is significantly improved. Furthermore, the patient flow shows that the RC fills the lungs significantly faster due to the improved pressure tracking. In Figure \ref{fig:pawenorm}, it is visible that the controller converges to a significantly smaller error 2-norm for all three scenarios. The improvement in error 2-norm is up to a factor 10 for the pediatric case. Furthermore, the controllers converge in approximately 5 breaths, which is considered sufficiently fast. 

Since the stabilizing controller used in addition to the RC is the same as benchmark controller, the same initial error 2-norm, i.e., during the first breath, is expected. However, another internal control loop is omitted from the RC, such that the RC takes care of this part as well. Therefore, a difference in initial error 2-norm is observed between the two control strategies. Furthermore, a significant difference is seen in error 2-norm from patient to patient. This is caused by the difference in breathing profile. A short breath, such as the baby breath, naturally results in a smaller error 2-norm, since the error 2-norm is not normalized to the breath length.


\section{Conclusions}\label{sec:Conc}
The presented repetitive control framework in this paper allows for superior pressure tracking performance for a wide variety of mechanically ventilated patients. This is achieved by using the periodicity of breaths in case a patient is fully sedated. To achieve this, a non-causal learning filter, based on an 'average patient', is designed. Thereafter, a robustness filter is designed to ensure stability. This robustness filter is designed such that the controller is robust with respect to plant variations. Finally, through experiments, it is shown that superior pressure tracking performance is achieved for a wide variety of patients.


\bibliography{PhDBibliography}

\end{document}